\documentstyle[11pt,lscape,amsmath,rotating]{article}

\def\be{\begin{equation}}
\def\ee{\end{equation}}
\def\bea{\begin{eqnarray}}
\def\eea{\end{eqnarray}}

\begin{document}

\title
{{\bf eRHIC - A precision electron-proton/ion collider facility at Brookhaven National Laboratory\footnote{Invited talk at the International Europhysics Conference on High-Energy Physics, Lisbon, Portugal, July 21-27, 2005.}}}

\author{B. Surrow \\
  {\it Massachusetts Institute of Technology, 77 Massachusetts Avenue, Cambridge, MA 02139} \\
  {\it Email: surrow@mit.edu}
}

\maketitle

\begin{abstract}
 An electron-proton/ion collider facility (eRHIC) is under consideration at Brookhaven 
 National Laboratory (BNL). 
 Such a new facility will require the design and construction of a new optimized detector
 profiting from the experience gained from the H1 and ZEUS detectors operated at the HERA collider at DESY.
 The details of the design will be closely coupled to the design of the interaction region, and thus to the 
 machine development work in general. An overview of the accelerator and detector design concepts will
 be provided.
\end{abstract}

\section{eRHIC accelerator design}
 The high energy, high intensity polarized electron/positron beam 
 ($5-10\,$GeV/$10\,$GeV) facility (eRHIC) which is under consideration at Brookhaven National Laboratory
 will collide with the existing RHIC heavy ion ($100\,$GeV per nucleon) and 
 polarized proton beam ($50-250\,$GeV). This facility will allow to  
 significantly enhance the exploration of fundamental aspects of Quantum Chromodynamics 
 (QCD), the underlying quantum field theory of strong interactions \cite{bs_deshpande,bs_jamal}. 
 A detailed report on the accelerator and interaction region (IR) design of this new collider facility
 has been completed based on studies performed jointly by BNL and MIT-Bates in collaboration with
 BINP and DESY \cite{1-ref}. The main design option is based on the construction of a $10\,$GeV electron/positron
 storage ring intersecting with one of the RHIC hadron beams. The electron beam energy will be variable down to
 $5\,$GeV with minimal loss in luminosity and polarization. The electron injector system will consist of linacs
 and recirculators fed by a polarized electron source. A study has shown that an ep luminosity of $4\times 10^{32}$cm$^{-2}$s$^{-1}$
 can be achieved for the high-energy mode ($10\,$GeV on $250\,$GeV), if the electron beam facility is designed using today's 
 state-of-the-art accelerator technology without an extensive R\&D program. For electron-gold ion collisions ($10\,$GeV on
 $100\,$GeV$/u$), the same design results in a luminosity of $4\times 10^{30}$cm$^{-2}$s$^{-1}$. The potential to go to higher 
 luminosities at the level of $10\times 10^{32}$cm$^{-2}$s$^{-1}$ (high-energy ep mode) by increasing the electron beam intensity 
 will be explored in the future. A polarized positron beam of $10\,$GeV energy and high intensity will also be possible using the
 process of self-polarization. A possible alternative design for eRHIC has been presented
 on the basis of an energy recovery superconducting linac (ERL) \cite{7-ref}. This option would be restricted to electrons only. 
 Preliminary estimates suggest that this design option could produce 
 higher luminosities at the level of $~10^{34}$cm$^{-2}$s$^{-1}$ (high-energy ep mode). Significant R\&D efforts for the 
 polarized electron source and for the energy recovery technology is required.

\begin{figure}[t]
\centering
\includegraphics*[width=120mm]{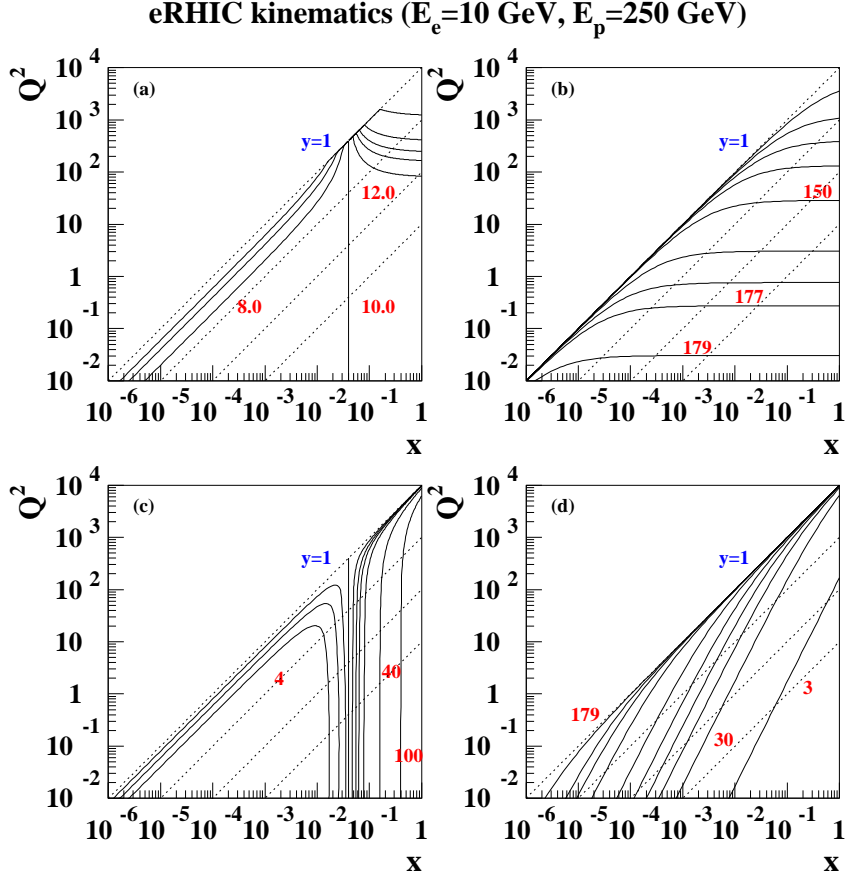}
\caption{\it Kinematic coverage for $10\,$GeV electrons on $250\,$GeV protons 
of the negative four-momentum squared $Q^{2}$ as a function of the Bjorken scaling variable $x$.
The kinematic coverage is shown for constant electron energy (a), constant electron scattering angle
(b), constant hadron energy (c) and constant hadron scattering angle (d). All angles are measured with
respect to the incoming proton beam. Hadron refers here to the hadronic final state associated to the
struck quark within the quark-parton model.}
\label{fig1}
\end{figure}

\section{eRHIC detector design}

The following minimal requirements on a future eRHIC detector can be made:

\begin{itemize}
\item Measure precisely the energy and angle of the scattered electron (Kinematics of DIS reaction)
\item Measure hadronic final state (Kinematics of DIS reaction, jet studies, flavor tagging, fragmentation studies, particle ID)
\item Missing transvere energy measurement (Events involving neutrinos in the final state, electro-weak physics)
\end{itemize}

\begin{figure}[t]
\centering
\includegraphics*[width=120mm]{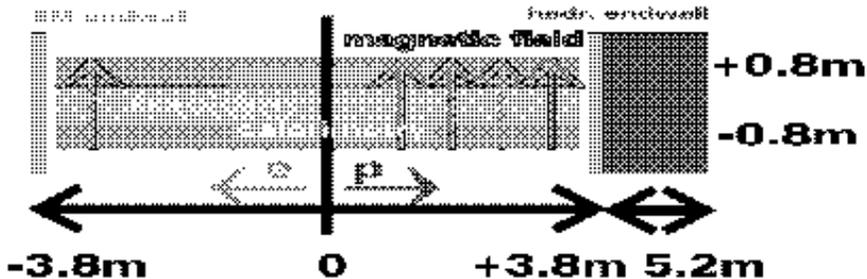}
\caption{\it Conceptual detector layout focusing on forward physics with a $7\,$m long dipole field and an
interaction region without machine elements extending from $-3.8\,$m to $+5.8\,$m.}
\label{fig2}
\end{figure}

In addition to those demands on a central detector, the following forward and rear detector systems are crucial:

\begin{itemize}
\item Zero-degree photon detector to control radiative corrections and measure Bremsstrahlung photons for luminosity measurements
\item Tag electrons under small angles (Study of the non-perturbative/perturbative QCD transition region and luminosity measurements from 
Bremsstrahlung ep events)
\item Tagging of forward particles (Diffraction and nuclear fragments)
\end{itemize}

Figure \ref{fig1} shows the kinematic coverage for $10\,$GeV electrons on $250\,$GeV protons 
of the negative four-momentum squared $Q^{2}$ as a function of the Bjorken scaling variable $x$.
The kinematic coverage is shown for constant electron energy (a), constant electron scattering angle
(b), constant hadron energy (c) and constant hadron scattering angle (d). All angles are measured with
respect to the incoming proton beam. Hadron refers here to the hadronic final state associated to the
struck quark within the quark-parton model.

Optimizing all the above requirements is a challenging task. Two detector concepts have been considered so far. 
One, which focuses on the forward acceptance and thus on low-$x$/high-$x$ physics which emerges out of HERA-III detector studies \cite{5-ref}. This detector concept is based on a compact system of tracking and central electromagnetic calorimetry inside a magnetic dipole field and
calorimetric end-walls outside. Forward produced charged particles are bent into the detector volume which extends the rapidity coverage compared to existing detectors. A side view of 
the detector arrangment is shown in Figure \ref{fig2}. The required machine element-free region amounts to roughly $\pm 5$m. This clearly limits the achieveable luminosity
in a ring-ring configuration. 

\begin{figure}[t]
\centering
\includegraphics*[width=120mm]{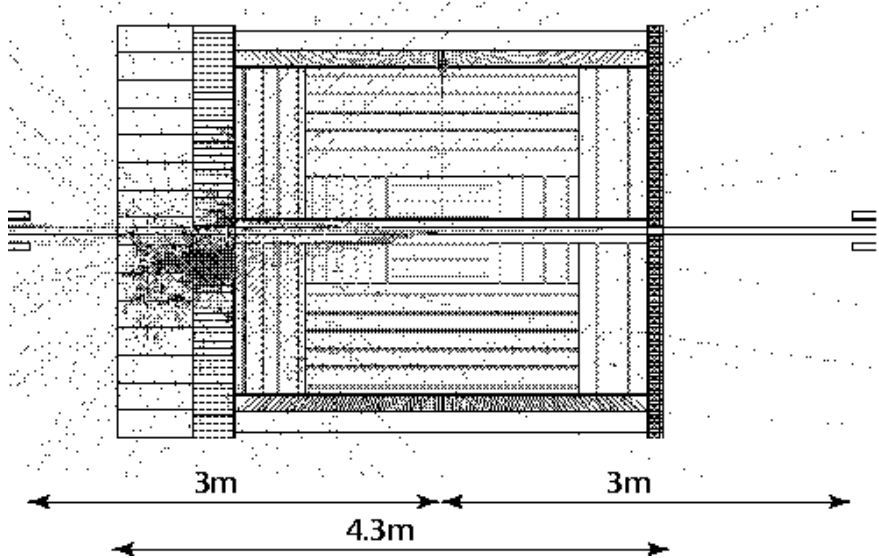}
\caption{\it Side view of the GEANT dector implementation as part of the ELECTRA simulation and reconstruction package \cite{4-ref}. 
A deep-inelastic scattering event resulting from a LEPTO simulation is overlayed with $Q^{2}=361\,$GeV$^{2}$ and $x=0.45$.}
\label{fig3}
\end{figure}

The second design effort focuses on a wide acceptance detector system similar to the current HERA collider experiments H1 and ZEUS to allow for
the maximum possible $Q^{2}$ range. The physics program demands high luminosity and thus focusing machine elements in a ring-ring configuration have to be as close as possible to the IR while preserving good central detector acceptance. This will be discussed in more detail in the next section. 
A simulation and reconstruction package called ELECTRA has been developed to design a new eRHIC detector at BNL \cite{4-ref,44-ref}. Figure \ref{fig3} shows a side view
of a GEANT detector implementation of the above requirements on a central detector. The hermetic inner and outer tracking system including the electromagnetic
section of the barrel calorimeter is surrounded by an axial magnetic field. The forward calorimeter is subidvided into hadronic and electromagnetic 
sections based on a conventional lead-scintillator type. The rear and barrel electromagnetic calorimeter consists of segmented towers, e.g. a tungsten-silicon
type. This would allow a fairly compact configuration. Other options based on a crystal rear and barrel electromagnetic calorimeter are under study.
The inner most double functioning dipole and quadrupole magnets are located at a distance of $\pm 3$m from the IR. An initial IR 
design assumed those inner most machine elements at $\pm 1$m. This would significantly impact the detector acceptance. More details on the IR
design can be found in \cite{1-ref}.

The bunch crossing frequency amounts to roughly $30\,$MHz. This sets stringent requirements on the high-rate capabililty of the tracking system. This makes
a silicon-type detector for the inner tracking system (forward and rear silicon disks together with several silicon barrel layers) together with several GEM-type outer 
tracking layers a potential choice. The forward and rear detector systems have not been considered so far. The design and location of those detector systems
has to be worked out in close collaboration to accelerator physicists since machine magnets will be potentially employed as sepectrometer magnetes and thus determine 
the actual detector acceptance and ultimately the final location. It is understood that demands on optimizing the rear/forward detector acceptance might have consequences
on the machine layout and is therefore an iterative process. 

\section{Considerations on the accelerator/detector interface}

The following section provides an overview of some aspects of the detector/machine interface. The specification of those items has only recently been started.

The direct synchrotron radiation has to pass through the entire IR
before hitting a rear absorber system. This requires that the geometry of the beam pipe is designed appropriately with changing shape along the longitudinal beam
direction which includes besides a simulation of the mechanical stress also the simulation of a cooling system of the inner beam pipe. The beam pipe design has
to include in addidition the requirement to maximize the detector acceptance in the rear and forward direction. Furthermore the amount of dead material has
to be minimized in particular to limit multiple scattering (track reconstruction) and energy loss for particles under shallow angles (energy reconstruction). The distribution
of backscattered synchrotron radiation into the acutal detector volume has to be carefully evaluated. An installation of a collimator system has to be worked out. Those
items have been started in close contact to previous experience at HERA \cite{6-ref}.

The demand of a high luminosity ep/eA collider facility requires the installaton of focusing machine elements
as close as possible to the central detector. An IR design with machine elements as close $\pm 1\,$m to the
IR which has been presented in \cite{1-ref} would significantly limit the achievable detector acceptance. A new
scheme has been presented in \cite{2-ref} which provides a machine-element free region of $\pm 3\,$m at the expense of approximatley
half the luminosity for the IR design presented in \cite{1-ref}. A linac-ring option would not be limited by beam-beam 
effects compared to a ring-ring configuration. Even larger luminosities could be achieved with a machine-element free region
of approximatley $\pm 5\,$m. This scheme has been presented in \cite{7-ref}.

The need for acceptance of scattered electrons beyond the central detector acceptance is driven by the need for luminosity
measurements through ep/eA Bremstrahlung and photo-production physics. Besides that a calorimeter setup to tag radiated photons 
from inital-state radiation and Brems- strahlung will be necessary. The scattered electrons will pass through the machine elements 
and leave the beam pipe through special exist windows. The simulation of various small-angle calorimeter setups has been started.
This will require a close collaboration with the eRHIC machine design efforts to aim for an optimal detector setup.

The forward tagging system beyond the central detector will play a crucial role in diffractive ep/eA physics. 
The design of a forward tagger system based on forward calorimetry and Roman pot stations is foreseen. Charged particles will be deflected
by forward machine elements. This effort will require as well a close collaboration with the eRHIC machine design efforts to ensure the best
possible forward detector acceptance.

\end{document}